\documentclass[aip,jcp,amsmath,amssymb,floatfix,reprint,twocolumn]{revtex4-2}
\usepackage{color,soul}
\usepackage{graphicx}
\usepackage{amsmath, amsthm, amssymb,mathtools}
\usepackage{amsthm}
\usepackage{multirow}
\usepackage[normalem]{ulem}
\usepackage{soul}
\usepackage{hyperref}

\usepackage{subfigure}

\usepackage{bm}
\usepackage{bigdelim}
\usepackage{array}
\usepackage{float}
\usepackage{subcaption}
\usepackage[table]{xcolor}
\arrayrulecolor{gray}
\newcommand{\mycomment}[1]{}

\newcommand{\be}{\begin{equation}}
\newcommand{\ee}{\end{equation}}
\newcommand{\bea}{\begin{eqnarray}}
\newcommand{\eea}{\end{eqnarray}}
\makeatletter

\newcommand{\Rmnum}[1]{\expandafter\@slowromancap\romannumeral #1@}
\makeatother\usepackage{array, makecell}

\begin{document}

\title{Inferring intermediate states by leveraging the many-body Arrhenius law}

\author{Vishwajeet Kumar}
\email{vishwajeet@imsc.res.in}
\affiliation{The Institute of Mathematical Sciences, C.I.T. Campus, Taramani, Chennai 600113, India}
\affiliation{Homi Bhabha National Institute, Training School Complex, Anushakti Nagar, Mumbai 400094, India}
\author{Arnab Pal} 
\email{arnabpal@imsc.res.in}
\affiliation{The Institute of Mathematical Sciences, C.I.T. Campus, Taramani, Chennai 600113, India}
\affiliation{Homi Bhabha National Institute, Training School Complex, Anushakti Nagar, Mumbai 400094, India}
\author{Ohad Shpielberg}
\email{ohads@sci.haifa.ac.il}
\affiliation{Department of Mathematics and Physics, University of Haifa at Oranim, Kiryat Tivon 3600600, Israel}
\affiliation{Haifa Research Center for Theoretical Physics and Astrophysics, University of Haifa, Abba Khoushy Ave 199, Haifa 3498838, Israel}


\begin{abstract}
 Metastable states appear as long-lived intermediate states in various natural transport phenomena which are governed by energy landscapes. As such, these intermediate metastable states dominate the system's dynamics at coarse grained times. Moreover, they can strongly influence the overall pathways through which the energy landscape is explored. Thus, quantifying these metastabilities is crucial for uncovering the key details of the underlying landscape.
Here, we introduce a robust method based on a generalized many-body Arrhenius law to identify metastable states in escape problems involving interacting particles with excluded volume.  Experimental platforms such as colloidal transport or macromolecular translocation through biological pores can offer promising settings to  validate our predictions.
\end{abstract}

\maketitle

\noindent

\section{Introduction}
The Arrhenius law (AL) is a cornerstone of physical chemistry, capturing the characteristic temperature dependence of reaction rates. It provides a quantitative framework for understanding diverse phenomena, from molecular reaction kinetics to everyday processes such as the faster spoilage of milk outside the refrigerator. 
It was the seminal work of Kramers, providing the underlying formalism that revealed the ubiquity of the AL throughout science and engineering \cite{hanggi1990reaction}. More than a century following Arrhenius’ experimental observation, the physics of activation still remains a vibrant topic of research \cite{woillez2019activated,woillez2020active,mukamel2005breaking,saadat2023lifetime,krapivsky2014large,berthier2011theoretical,langer1969statistical, dubkov2023enhancement,del2023escape,neupane2016direct,chupeau2020optimizing,yu2012direct,thorneywork2020direct,thorneywork2024resolution,ferrer2024experimental,ginot2022barrier,lapolla2020single,kumar2025speeding}.
In its simplest form, the Kramers problem consists of an overdamped particle in a trapping potential, coupled to thermal bath at temperature $T$. To escape the trapping potential, the particle needs to overcome the energy barrier denoted by $\Delta U$ - the activation energy. In the limit of deep traps $\beta \Delta U \gg 1$, the single particle escape rate is captured by the AL
\begin{equation}
\label{eq:AL sp}
\Phi_\text{SP} =  \frac{1}{\tau_0} \rm{e}^{-\beta \Delta U}, 
\end{equation}
where, $\beta=1/k_BT$ with $k_B$ being the Boltzmann constant. An important feature of the escape rate given in Eq.~\eqref{eq:AL sp} is that the exponential factor is universal, depending solely on the barrier height~$\Delta U$, and thus remains invariant under changes in the detailed shape of the potential. In contrast, the prefactor~$\tau_0$ represents a non-universal microscopic time scale which crucially depends on the specific form of the potential landscape\cite{hanggi1990reaction,gardiner1985handbook,elber2020molecular}.  

Of prime importance is the inverse problem: extracting the features of the energy landscape or more precisely, the details of the potential trap from the escape time statistics. One aspect of the inverse problem, prevalent in the theory of chemical dynamics, is predicting the value of the energy barrier $\Delta U$ by interpreting experimental data. Such an activation process may extend beyond the standard Kramers setup, and involve a conformational or configurational transition as motion along the reaction coordinate in a rugged energy landscape \cite{makarov2015single}.  
A second aspect of the inverse problem is inferring the coordinates and number of local minima in the energy landscape. The local minima can be identified as long-lived states, or metastabilities, where the system spends  significant time.  
Identifying these metastabilities enables the reduction of the system’s state space by filtering out fast degrees of freedom, a crucial technique for simplifying complex dynamics in chemical physics. While the AL Eq. \eqref{eq:AL sp} is equipped to address finding $\Delta U$ as the slope of the semilog plot of $\Phi_{\rm{SP}}$, building on the subexponential dependence on $\tau_0$, it is ill equipped to infer the number and positions of the metastabilities.

The problem of quantifying the state space of long-lived states was addressed in the literature, especially by employing statistical tools to assess the state space number, i.e. the number of local minima of the trapping potential or metastabilities.  
For instance, one key result in chemical kinetics, is that the randomness parameter or the coefficient of variation measured from the turnover time of an enzymatic reaction can place a strict lower limit on the topology i.e., the number of  metastable kinetic states in any viable model for the mechanism of a given enzyme \cite{moffitt2010methods}. 
Another method based on escape time was proposed by Li and Kolomeisky \cite{li2013mechanisms} to determine the mechanism and topology of complex chemical and biological networks. There, the authors hypothesized that the dwell-time distributions of events between two
states can encode information on number of intermediate states, pathways, and transitions that lie between initial and final states.
Recently, this theoretical work was tested in a high precision experiment \cite{thorneywork2020direct}, for an overarching set of complex molecular or nanoscale systems to unravel the quantitative features of the underlying energy landscape. 
In another work, the structure of the protein folding energy landscape involving multiple kinetic traps was predicted via the multiple relaxation time scales emanating from the first passage time distribution between reactants and products of these biomolecules  
\cite{wales2022dynamical}.  In a similar vein, here we show that the generalized Arhhenius law, Eq. \eqref{eq:gen AL}, enables the inference of the number of metastabilities in a potential landscape--- thereby establishing a powerful and robust theoretical 
method irrespective of the complex underlying mechanisms, and which could not be achieved within the traditional Arrhenius law, Eq. \eqref{eq:AL sp}.

\begin{figure*}
    \centering
    \includegraphics[width=\linewidth]{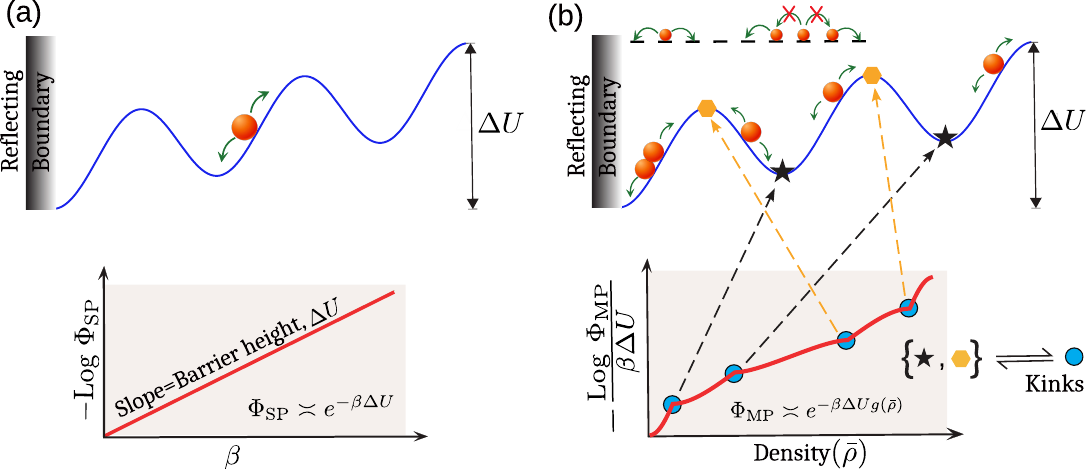}
    \caption{(a) The escape problem of a single particle. The bottom figure shows the traditional inference of the activation barrier height from Arrhenius law Eq. \eqref{eq:AL sp}  (b) Inference of the potential minima and maxima by employing the many-body Arrhenius law Eq. \eqref{eq:gen AL}. The number of kinks in the bottom figure corresponds to local maxima and minima of the potential $U(x)$ in the top figure. Thus, the many-body escape dynamics 
    (for instance, activated dynamics of a stochastic lattice gas model with volume exclusion as shown in the top)
    allow to infer the metastabilities of the single particle escape problem. 
    }
    \label{fig:main_figure}
\end{figure*}

Diffusive particles with excluded volume interactions, positioned in a deep trap, acquire the escape rate given by the many-body Arrhenius law \cite{kumar2024arrhenius,kumar2024emerging} 
\begin{equation}
\label{eq:gen AL}
    \Phi_\text{MP} \asymp \exp\left[-  \beta \Delta U g(\bar{\rho}) \right] , 
\end{equation}
where $\Delta U g(\bar{\rho})$ is an effective activation energy \footnote{Here $\asymp$ suggests asymptotic equivalence, discarding sub-exponential corrections.}. 
The function $g(\bar{\rho})$ represents the many-body correction to the conventional Arrhenius law. Notably, while $g(\bar{\rho})$ depends on the particle density $\bar{\rho}$ and on the specific form of the trapping potential, it remains independent of the microscopic details of the particle interactions. The only requirement is that the interactions include, but are not restricted to hard core repulsion -- there exist an excluded volume (see \cite{kumar2024emerging} for specific examples).  
Moreover, the function $g(\bar{\rho})$ is computable for an arbitrary trapping potential, using a geometric argument. 
In this work, we show that for a given potential, varying $\overline{\rho}$ --  the particle density in the trap may lead to kinks in $g(\overline{\rho})$. These kinks are associated with the local  minima and maxima of the trap potential (see FIG.~\ref{fig:main_figure}).   
  { }We reveal a surprising equivalency between $\Phi_{\rm{MP}}$ and a thermodynamic partition function. Drawing on the theory of critical phenomena, we construct analogous thermodynamic response functions as quantitative tools to detect metastabilities, treating them as the counterparts of thermodynamic phase transitions.

\section{Setup}

To make matters more concrete, let us introduce a setup \`a la Kramers in one dimension. 
  We confine an interacting gas consisting of $M$ excluded volume particles in a trap of size  $ L$, with $X$ denoting the position in the trap. The gas particles can exhibit non-trivial short-ranged interactions, beyond the excluded volume, as long as the resulting dynamics possess diffusive dynamics. We rescale the position such that $x=X/L$ and the rescaled trapping potential is denoted by $U(x)$, where $x\in[0,1]$ is dimensionless. As a rule, we keep to dimensionless units in this work.  
 The trap potential has a global minimum at $x=0$. We fix $U(0)=0$ for convenience, and impose a reflecting boundary condition at $x=0$.  At $x=1$, the potential has a global maximum, fixing $U(1)=\Delta U$, and imposing an absorbing boundary condition.  
 We are interested in the mean escape time, i.e the mean time it takes for the fastest particle to reach the absorbing boundary at $X=L$ for the first time. In this work, we focus on deep potential traps, i.e. $\beta\Delta U \gg 1$,  where the mean escape time is long, and as large deviation theory becomes relevant, the mean escape time converges to $1/\Phi_{\rm{MP}}$ , under mild assumptions on the initial conditions (to be discussed later on).

\begin{figure}
    \centering
    \includegraphics[width=\linewidth]{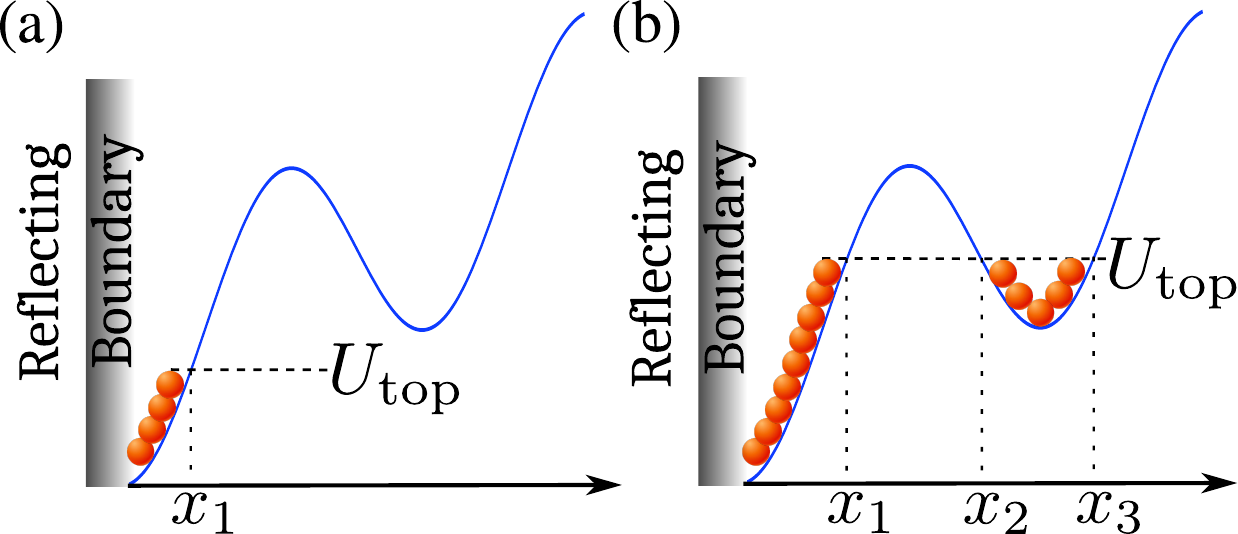}
    \caption{A visualization of the MEC for particles with excluded volume on an arbitrary non-monotonic potential. Particles occupy the potential for $U(x)<U_{\rm{top}}$. (a) Here $U_{\rm{top}}$ is smaller then the lowest value of the potential extrema. Thus, $U_{\rm{top}} $ has a single intersection with $U(x)$, at position $x_1$, and we find $\overline{\rho}(U_{\rm{top}} = U(x_1) ) $, leading to $U_{\rm{top}}(\overline{\rho}) =  U(x=\overline{\rho})$ as in the monotonous potential. 
    (b) Here $U_{\rm{top}} $ is in the range where there are three intersection points with $U(x)$, denoted by $x_{1,2,3}$. In this case  $\overline{\rho}(U_{\rm{top}}) = x_1 + x_3-x_2$. Notice that in this case $U_{\rm{top}}(\overline{\rho}) \neq  U(x=\overline{\rho})$. 
    }
    \label{fig:MEC}
\end{figure}

 In the limit of large $M,L$, with a fixed particle density in the trap $\overline{\rho}=M/L$ \cite{kumar2024emerging,hartich2018duality}, we find  Eq. \eqref{eq:gen AL}. The form of $g$ in Eq. \eqref{eq:gen AL} is explicitly given by $g(\bar{\rho}) = 1- U_{\rm{top}}(\bar{\rho})/\Delta U$, with 
 $U_{\rm{top}}$ defining the maximum potential value occupied by particles  when they are arranged in the minimum energy configuration (MEC) (see FIG.~\ref{fig:MEC}). This implies  
 \begin{equation}
 \label{eq:heaviside rho}
 \overline{\rho} = \int dx \,  \Theta (\textcolor{black}{\beta} U_{\rm{top}} - \textcolor{black}{\beta} U(x)),  
 \end{equation}
  where $\Theta $ is the Heaviside function (see the appendix B for connection with microscopic details). 

 Eq. \eqref{eq:gen AL} is non-trivial to derive in the many-body case, yet, it can be understood by an intuitive argument. 
  In the limit $\beta \Delta U \rightarrow \infty$, the particles organize into the  MEC, which is primarily determined by the structure of the potential (see FIG.~\ref{fig:MEC}). The escaping particle is the one with the highest potential energy, $U_{\rm{top}}$ in the MEC ( also see Langer et al.\cite{langer1969statistical}).
This intuition essentially leads to  Eq. \eqref{eq:gen AL}, recalling that $g(\bar{\rho})=1-U_{\rm{top}}(\bar{\rho})/\Delta U$. Note that activating a single particle from the MEC leads to an energy dominated free energy change in the macroscopic system. The latter assertion may differ for particles with long-range interactions like polymers \cite{sebastian2000kramers} which we do not consider within our formalism. 

For the sake of concreteness, we stress that our analysis relies on two assumptions. First, that the typical time scale for activation is much larger than the diffusive time scales. Second, that the activation time from the initial condition is much longer than the relaxation time to the MEC. Physically, this means that there is no set of particles that start too close to the absorbing boundary, or a macroscopic set of particles that starts in a local potential minimum, away from the origin, and detached from the main bulk of particles.

The MEC was shown to correctly predict $g(\bar{\rho})$ for a set of important diffusive lattice gas models using a hydrodynamic theory \cite{kumar2024arrhenius,kumar2024emerging}, where the potential was taken to be monotonous. Indeed, it has already been established that hydrodynamic theories can correctly predict current statistics in some special cases \cite{agranov2016survival,agranov2018narrow,krapivsky2014large,grabsch2024semi}.  

Within our trap setup, and for monotonous potentials, the MEC predicts via Eq. \eqref{eq:heaviside rho}  that $U_{\rm{top}} = U(x=\overline{\rho})$, leading to 
\begin{equation}
\label{eq:monotone g}
    g_{\textrm{monotone}}(\overline{\rho}) = 1 - U(x=\overline{\rho})/\Delta U.
    \end{equation}
    See also FIG.~\ref{fig:MEC}.  
    Intuitively, the escape rate should increase as we fill the trap with excluded volume particles, since adding particles pushes the highest energy particle in the trap closer to the activation energy. Hence, $g(\bar{\rho})$ has to be a monotonous decreasing function of $\overline{\rho}$. From this reasoning alone, it is clear that $g(\bar{\rho})$ cannot follow Eq. \eqref{eq:monotone g} for non-monotonous potentials.  
    The presence of non-monotonic potentials is quite ubiquitous in the context of particle, ion or metabolite transport through confined geometries containing narrow openings, corners, bottlenecks or channels. Despite their pervasiveness, the inverse problem of inferring the structure of the channel, the nature of the potential from the escape time statistics remains a challenge \cite{wales2022dynamical}. In what follows we analyze the structure of $g(\bar{\rho})$ for an arbitrary non-monotonic potential, and show the relevance to the inference problem.

\section{Emergence of the kinks}
In non-monotonic potentials, the escape rate, and in particular $g(\overline{\rho})$ varies dramatically from the monotonic case captured in Eq. \eqref{eq:monotone g}. Indeed, one still infers $g(\bar{\rho})$ from the MEC. Before providing $g(\bar{\rho})$ for an arbitrary trapping potential, it is instructive to first consider a minimal model for a non-monotonic potential -- the piecewise linear potential (see FIG.~\ref{fig:Upwl and 1-g}). 
Recall that $g(\bar{\rho})=1-U_{\rm{top}}(\bar{\rho})/ \Delta U$ for any arbitrary potential. However, crucially $U_{\rm{top}} \neq U(x=\overline{\rho})$ for the non-monotonous potentials. 
Now, one may proceed by determining 
$U_{\rm{top}}(\overline{\rho})$, or alternatively the inverted function $\overline{\rho}(U_{\rm{top}})$ from Eq. \eqref{eq:heaviside rho},
the latter leads to 
\begin{equation}
\label{eq:piecewise U of rho}
    \overline{\rho} (U_{\rm{top}}) = 
\begin{cases}
\frac{U_{\rm{top}}}{2\Delta U} & \text{for } 0<\frac{U_{\rm{top}}}{\Delta U}<\frac{1}{3}  \\
2\frac{U_{\rm{top}}}{\Delta U}-\frac{1}{2}, & \text{for } \frac{1}{3}<\frac{U_{\rm{top}}}{\Delta U}<\frac{2}{3}\\
\frac{\frac{U_{\rm{top}}}{\Delta U}+1}{2}, & \text{for } \frac{2}{3}<\frac{U_{\rm{top}}}{\Delta U}<1.
\end{cases}
\end{equation}

It should then be understood that the kinks in $\overline{\rho} (U_{\rm{top}})$ correspond to the local minimum and maximum of $U_{\rm{pwl}}$. These kinks would be inherited by $g(\overline{\rho})$. Recalling $g(\overline{\rho})=1-U_{\rm{top}}/\Delta U$, one can invert Eq. \eqref{eq:piecewise U of rho}, to find (see FIG.~\ref{fig:Upwl and 1-g}) 
\begin{equation}
\label{eq:g for pwl potential}
    1 - g (\overline{\rho}) = 
\begin{cases}
2 \overline{\rho} & \text{for } 0<\overline{\rho}<\frac{1}{6}  \\
    \frac{1}{2}\overline{\rho} + \frac{1}{4} & \text{for } \frac{1}{6}<\overline{\rho}<\frac{5}{6}\\
2\overline{\rho}-1 & \text{for } \frac{5}{6}<\overline{\rho}<1
\end{cases}
\end{equation}

The treatment above for $U_{\rm{pwl}}$ can be generalized to an arbitrary potential trap. It is the local minima and maxima of the potential trap that introduce kinks in $g(\overline{\rho})$.  
Let $n$ be the number of intersections of $U_{\rm{top}}$ with an arbitrary trapping potential $U(x)$. Unless $U_{\rm{top}}$  exactly intersects with a local maximum or minimum point, $n=2m+1$ is an odd number larger or equal to one. The MEC implies that
\begin{equation}
\label{eq:gen rho of Utop}
\overline{\rho}(U_{\rm{top}}) = \sum^{m} _{i=0} \left( x_{2i+1} - x_{2i}\right),    
\end{equation}
 where $x_0 =0$ and $x_i$ is the $i$-th intersection of $U(x) $ with $U_{\rm{top}}$  (FIG.~\ref{fig:MEC} (b) demonstrates the idea). The origin of the kinks observed in $g(\bar{\rho})$ can be understood as follows. See Fig.\ref{fig:main_figure}(b). For small values of $\bar{\rho}$, the particles arrange themselves around the reflecting boundary in the MEC. As $\bar{\rho}$ increases further beyond a certain value, the MEC paradigm suggests that the particles begin to populate not only the region near the reflecting boundary but also the vicinity of the first potential minimum (which lies at a lower energy level than the second minimum; see FIG.~\ref{fig:main_figure}(b)). This shift in spatial distribution of particles marks the emergence of the first kink in $g(\bar{\rho})$ shown in FIG.~\ref{fig:main_figure}. With subsequent increases in density, similar organizational shifts occur, leading to additional kinks in $g(\bar{\rho})$. The number of kinks equals the number of local minima and maxima of the potential $U$ as shown in FIG. \ref{fig:main_figure}(b). 
It should be noted that it is hard in general to provide an analytic expression of $g(\overline{\rho})$ as it involves finding analytically the intersections of the arbitrary trap potential $U(x) = U_{\rm{top}}$ for any  $0<U_{\rm{top}}<1$, and then invert Eq. \eqref{eq:gen rho of Utop}. Nevertheless, a numerical inference of $g(\bar{\rho})$ is straight-forward for any arbitrary trap potential and at any desired level of precision. Thus, we are able to represent $g(\bar{\rho})$ for both monotonous and non-monotonous trap potentials using the MEC. 

\section{The kinks and thermodynamic phase transitions}
Importantly, the MEC and the resulting kinks are obtained as asymptotic results in the  $\beta \Delta U \rightarrow \infty$ limit. The kinks constitute dynamical phase transitions as the mean density is varied. However, the escape rate $\Phi_{\rm{MP}}$, or $g(\bar{\rho})$, are expected to be analytic at finite $\beta \Delta U$. To build this intuition, let us draw a correspondence between the kinks in the escape problem to the critical phenomena in thermodynamics. A thermodynamic phase transition points to a non-analyticity in the thermodynamic potential, say the Helmholtz Free energy, when the temperature is varied beyond a critical temperature. The non-analytic point is apparent only in the thermodynamic limit of an infinite system. Translating to the MEC, the \textit{large system size} in thermodynamics corresponds to the $\beta \Delta U \rightarrow \infty$ limit and the \textit{temperature}  corresponds to the mean density $\overline{\rho}$,  which can be externally controlled. Bridging this correspondence, we can define 
\begin{equation}
\mathcal{F}=- \frac{1}{\beta \Delta U}\log \Phi_{\rm{MP}}    , \quad 
\mathcal{R}_n = \frac{\partial^n \mathcal{F} } {\partial {\overline{\rho}}^n}
\label{Free-energy-PT}
\end{equation}
 to correspond to the free energy per volume, where $\beta\Delta U $ serves as the volume parameter and $\mathcal{R}_n$ are the associated response functions.  It is important to note that in the limit $\beta \Delta U\to\infty$, the free energy density $\mathcal{F}$ corresponds to $g(\bar{\rho})$, and that the singular behavior arises from $\Phi_{\rm{MP}}$ which corresponds to a partition function in statistical mechanics.

Crucially, the non-analytic nature of the escape rate becomes apparent only in the \textit{thermodynamic limit}, i.e. at $\beta \Delta U \rightarrow \infty $. At finite $\beta \Delta U$, it is especially hard to observe the emerging kinks, as  $\beta\Delta U$ is rather limited experimentally to order of $20$, contrary to the macroscopic volume in thermodynamic systems (see Appendix \ref{sec:More models}, FIG.\ref{fig:TE potential Numerics} and FIG.\ref{fig:Uxs potential }). 
A standard method to observe phase transition is by measuring the scaling of response functions such as susceptibilities and other derivatives of the thermodynamic potentials. This essentially corresponds to looking at the derivatives of $\mathcal{F}$ and the scaling of the divergences close to the critical densities, i.e. where we expect the kinks. That is, instead of the MEC, it is required to determine $\Phi_{\rm{MP}}$ for a finite $\beta \Delta U$. This has been achieved \cite{kumar2024emerging}, by developing a perturbative approach, estimating $g$ analytically at a finite (but large) $\beta\Delta U$ \footnote{In \cite{kumar2024arrhenius}, an exact approach was developed. However, the treatment is much harder analytically, and numerically the estimation is limited to about $\beta \Delta U = 20$. }. From the knowledge of $\Phi_{\text{MP}}$ at finite $\beta \Delta U$, the corresponding finite-$\beta \Delta U$ response functions can be obtained using Eq.~\eqref{Free-energy-PT}. These response functions capture the signatures of kinks in $g$, thereby revealing the number of metastable states in the potential landscape.

Statistical physics states that   finite systems can only lead to analytic thermodynamic observables. Therefore, phase transitions occur only in the thermodynamic limit of infinite system size. However, for a macroscopic particle number of the order of Avogadro's number, it is hard to tell the difference between a true divergence of, e.g. the response functions, and a large but finite increase close to the critical point. Here, $\beta \Delta U $ serves as the equivalent of the system size, whereas the density serves as the control parameter to manipulate towards the critical points. This implies that observing the kinks experimentally is unattainable. It is more practical to focus on the scaling properties of the diverging response functions. To see that, we recall that 
\begin{align}
\label{eq:Z scaling}
    \mathcal{F} &=  g(\overline{\rho}) + \mathcal{F}_{\rm{s}}
     + \textrm{non-singular subleading terms} ,
      \nonumber \\
     \mathcal{F}_{\rm{s}} &=  \Delta U^{-\alpha} f_{\rm{s}}\left(\beta\Delta U \delta \rho^\gamma \right).
\end{align}
The regular part is dominated by $g(\bar{\rho})$ at large $\beta \Delta U$, and the singular part is given by a singular scaling function $f_{\rm{s}}$. Here $\delta \rho$ is the distance from the critical density, and $\alpha,\gamma >0$ are critical exponents.  We note, however, that the inference of kinks in $g(\bar{\rho})$ doesn't require the exact determination of these critical exponents.

\section{
Response functions for an interacting lattice gas model}
 
To demonstrate the emergence of the kinks through the response functions, we focus on the simple exclusion process (SEP) \cite{mallick2015exclusion,derrida2007non,lips2020nonequilibrium,lips2019single}. 
The SEP is a paradigm in the study of nonequilibrium physics,  capturing the behavior of stochastic transport of biomolecules and molecular motors through complex media, surface growth, vehicular traffic and more  \cite{mallick2015exclusion,chou2011non,appert2011frozen}.

The SEP describes a jump process, where particles jump to neighboring empty lattice sites only thus ensuring an excluded volume interaction. 
To find the finite size $\beta \Delta U$ escape rate,  Kumar et al. \cite{kumar2024emerging} have used a hydrodynamic theory  -- the macroscopic fluctuation theory \cite{bertini2015macroscopic,agranov2018narrow,shpielberg2018universality} --  to handle the inter-particle interactions. The theory shows that one needs to find the optimal density profile that maximizes the probability that all the particles survive in the trap. Here, there is underlying assumption, previously stated, that the initial density profile can relax to this optimal density profile in time scale much shorter than the activation time.  
To that end, it was proposed   to consider a perturbative approach \cite{kumar2024emerging}. First, consider the equilibrium density profile in the box with reflecting boundary conditions. In this case, the density at the right hand side does not vanish as it should for an absorbing boundary condition. However, the equilibrium density at the absorbing boundary is small due to the high value of the potential $\beta U(x=1)=\beta\Delta U$. So, the typical density profile can be obtained by adding a small perturbation to the equilibrium density profile, which takes care of the absorbing boundary discrepancy. 

\begin{figure}[t!]
    \centering
\includegraphics[width=\linewidth]{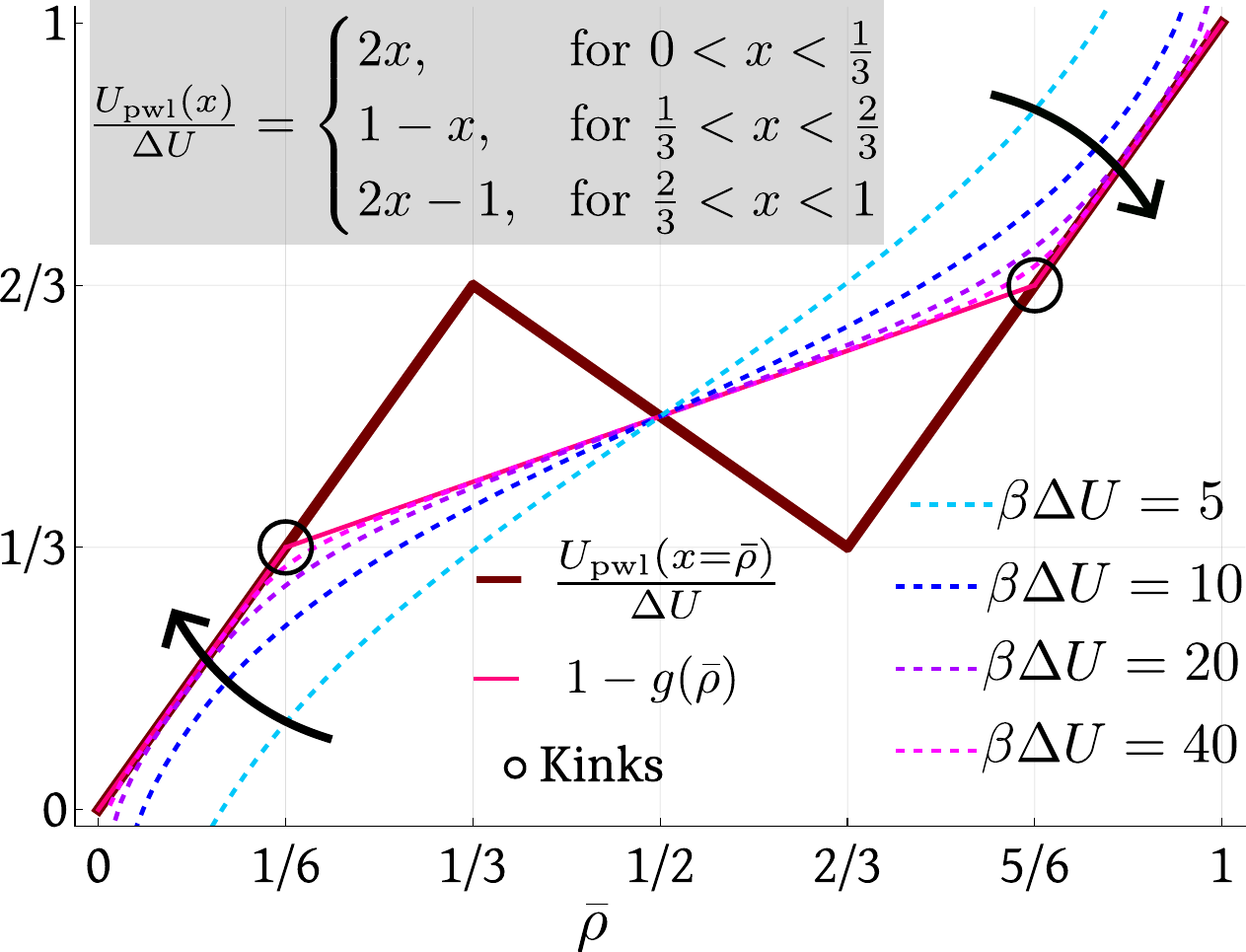}
  \caption{(Top) The piecewise linear potential $U_{\rm{pwl}}(x)$ is plotted (solid thick brown) together with the MEC prediction for $1-g$ (solid magenta). The black circles highlight the positions of the kinks in $g$. The dashed lines reveal the convergence of the finite $\beta \Delta U$ approximation scheme to the MEC $1-g$ curve. Noticeably, the convergence, marked by the black arrows, is faster away from the kinks.}
 \label{fig:Upwl and 1-g}
 \end{figure}

We start with the general expression for the escape rate namely $\Phi_{\rm{MP}} \asymp A \rm{e}^{-\beta \Delta U} $, as derived in the reference\cite{kumar2024emerging}. Here, one can infer $A$ from the value of the equilibrium density profile at the absorbing boundary using the following relation (see Appendix \ref{sec:MFT}, \ref{sec:survival mft} and \ref{sec:SEP Upwl analysis})
\begin{equation}
\label{eq:Formal A-0}
    1- \overline{\rho} = \int dx \, \frac{1}{1+A \rm{e}^{-\beta U(x)}},
\end{equation}
for a given $U(x)$. For the SEP particles in a piecewise linear potential $U_{\rm{pwl}}$, ~Eq. \eqref{eq:Formal A-0} can be solved to get an expression for $A$ in the ``finite size'' limit,
revealing the precursor to the criticality in the escape rate in the ``finite size'' limit (see Appendix \ref{sec:SEP Upwl analysis}).  This determines $\Phi_\text{MP}$ at finite $\beta\Delta U$. Subsequently, the corresponding free energy density $\mathcal{F}$ is computed from $\Phi_{\text{MP}}$ using Eq. \eqref{Free-energy-PT} and is shown in FIG. \ref{fig:Upwl and 1-g} as dashed curves for various $\beta\Delta U$ values.
We next examine in greater detail  the regime $0<\overline{\rho}<1/6$ where the finite size expression of $A$ (and thus the escape rate) can be made simpler (see Appendix \ref{sec:SEP Upwl analysis}). 
However, the analysis could be extended to the full range of $\overline{\rho}$, doing so in the main text would introduce  technical complexity, going against our pedagogical purposes. Thus, we provide a full numerical analysis of the piecewise linear potential in Appendix \ref{sec:SEP Upwl analysis}.

\begin{figure}
    \centering
    \includegraphics[width=\linewidth]{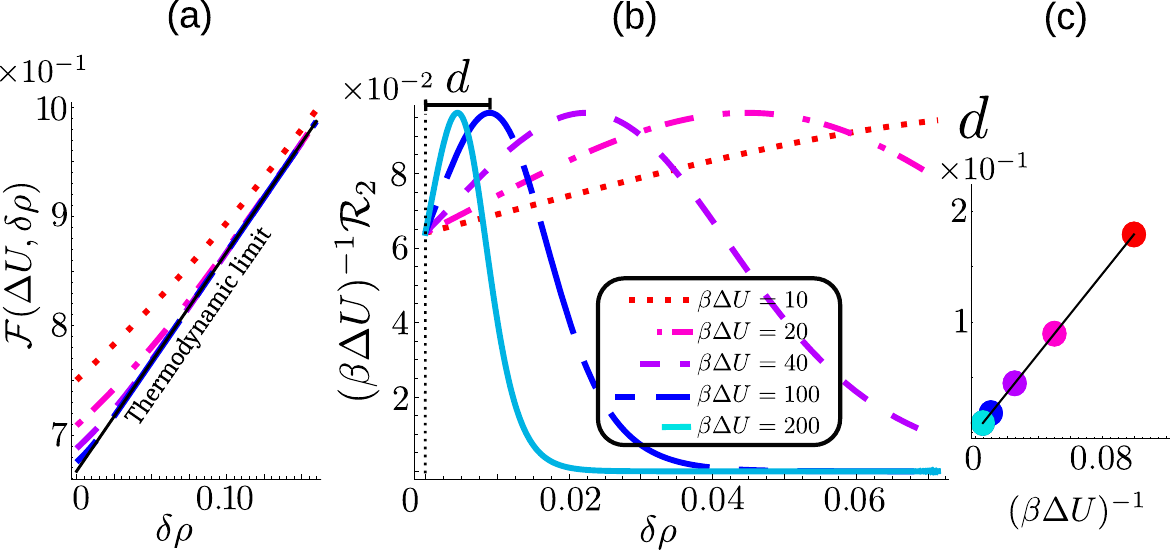}
    \caption{ (a) The ``free energy'' $\mathcal{F}$ is calculated for  finite  $\beta \Delta U$ values, and compared with the MEC result ($\beta \Delta U=\infty$) for SEP particles subjected to the $U_{\rm{pwl}}$ trapping potential. It is evident that the convergence is better away from the critical point at $\delta \rho =0$. However, to infer the criticality, we need to consider the response function $\mathcal{R}_2$. (b) The rescaled response function $\mathcal{R}_2/(\beta\Delta U)$ for different $\beta \Delta U$ as found from ~Eq. \eqref{eq:F analytic}. The precursor of the kink manifest as a scalable peak, at distance $1/ \beta \Delta U$ from the expected kink. (c) The distance $d$ of the peaks from the expected critical point is shown to scale like $1/\beta \Delta U$. 
    We remind that in finite systems, the singularities characteristic of critical phenomena are rounded and shifted: response functions exhibit smooth peaks rather than divergences. As the system size increases, these peaks grow sharper and their location approaches the true critical point in the thermodynamic limit. Here, we observe the same for $d$ as $\beta \Delta U$ replaces the system size \cite{kardar2007statistical}.  
    }
    \label{fig:piecewise potential scaling}
\end{figure}

 We can now write the corresponding free-energy in the finite size limit following  (\ref{Free-energy-PT})
\begin{eqnarray}
\label{eq:F analytic}
    \mathcal{F} &=& \mathcal{F}_0 + 
    \mathcal{F}_{\rm{s} } , \\ \nonumber 
    \mathcal{F}_{\rm{s} }  &=& -\frac{1}{\beta\Delta U}\log \frac{ -1 + \sqrt{1+12\rm{e}^{-2\delta \rho  \beta\Delta U}}}{6} , 
\end{eqnarray}
where $ \delta \rho =1/6- \overline{\rho} $ measuring the distance from the critical density $\overline{\rho} =1/6$. Here $\mathcal{F}_0  = 2/3$ is a regular contribution that will not lead to the divergence of the response functions. We note that there are subleading terms in $\mathcal{F}_0$, that are discarded in our perturbative approach. $\mathcal{F}_{\rm{s} } $ is the singular part of the associated free energy. The singularity appears at $\delta \rho  \beta \Delta U \rightarrow \infty$ which corresponds to a finite $\delta \rho $ (distance from the critical density) and infinite $\beta \Delta U$.  It is then clear that observing the singularity directly from $\mathcal{F}$ is impractical.   
In other words, 
the formation of the kinks will become apparent only at exceptionally large values of $\beta \Delta U \approx 100$. Therefore, to observe the precursors of the kinks, we turn to the response functions. The response functions accentuate the singular part of $\mathcal{F}$, i.e. $\mathcal{R}_n \propto (\beta \Delta U)^{n-1}$ when keeping $\beta \Delta U \delta \rho$ finite.  FIG.~\ref{fig:piecewise potential scaling} demonstrates our findings. We first show that while $\mathcal{F}$ converges to its MEC value as we increase $\beta \Delta U$, the convergence is slower close to the critical point, i.e. for $\delta \rho  \rightarrow 0$. Moreover, we demonstrate the scaling behavior of the response function $\mathcal{R}_2$. 

Finding an analytical expression for $\mathcal{F}$ for arbitrary potential landscape turns out to be  completely intractable, even for the SEP within the perturbative approach. Nevertheless, inference of the kinks through the response functions is not limited to an analytic treatment alone, but can be probed using numerical techniques using the response functions.  
This is demonstrated in Appendix \ref{sec:More models}
for nonlinear trap potentials  
with multiple metastabilities -- the appearance of similar potential energy barriers in the activation gating of the bacterial ion channels is quite ubiquitous \cite{linder2013probing}. 
It is demonstrated in Appendix (see FIG. \ref{fig:TE potential Numerics} and FIG. \ref{fig:Uxs potential }) how $\mathcal{F}$ systematically converges to the MEC limit in the large $\beta \Delta U$ limit while the response functions clearly indicate the emergence of the kinks. In a nutshell, the local minima and maxima of the potential can be inferred from the response functions, obtained from escape time measurements, without prior knowledge of the trap potential. 

\section{Discussion}

In this work, we have revisited the survival probability of interacting particles located in a deep trap potential. The particle escape rate follows the generalized Arrhenius form $\Phi_{\rm{MP}} \asymp \rm{e}^{-\beta \Delta U g(\bar{\rho}) }$, where $g(\bar{\rho})$ can be asymptotically assessed via the MEC. It was shown that in the excluded volume universality, i.e. for lattice gas models where the lattice site occupancy is capped, $g(\overline{\rho})$ becomes singular in $\beta \Delta U \rightarrow \infty$ when the trap potential introduces local minima and maxima. More precisely, local maxima and minima of the trap potential lead to kinks in  $g (\overline{\rho})$. Thus, reversing the logic, measurement of kinks in the many-body $g(\bar{\rho})$ allows to infer the number of the single particle long-lived states (metastabilities) induced by the trap potential local minima. The salient point of this work is proposing a statistical test, based on the measureable mean escape time of the many-body problem, that allows to quantify the number of  long-lived states in the single-particle activation problem.

 By establishing a correspondence between the kinks in $g(\bar{\rho})$ in the escape problem and thermodynamic phase transitions, we introduced the free energy $\mathcal{F}$ and the associated response functions $\mathcal{R}_n$ in Eq.~\eqref{Free-energy-PT}. The response functions $\mathcal{R}_n$ (for $n \ge 2$) capture the signature of emergence of kinks at finite $\beta \Delta U$ values, which would otherwise manifest only in the asymptotic limit $\beta \Delta U \to \infty$. In the example considered here, we found  that the critical exponents related to the scaling functions (see Eq.~\eqref{eq:Z scaling}) are  $\alpha=\gamma =1$. However, we emphasize  that 
determining the exact values of these exponents is not essential (and 
lies beyond the scope of the present study), identifying the onset of them is sufficient for our problem. Thus, our key objective has been to develop a framework that can identify the criticality, rather than its precise nature. Notably, one might be tempted to claim universality of the critical exponents due to symmetries of the Lagrangian. However, a recent work \cite{schorlepp2025systematic} by one of us suggests that while the critical exponents are indeed universal for smooth potentials, they can be different for non-differentiable piecewise linear potentials.  
This is indeed what we have observed numerically (see captions of Figs.~\ref{fig:Upwl Full Numerics},~\ref{fig:TE potential Numerics} and \ref{fig:Uxs potential }).

As a proof of concept, it was demonstrated that this approach allows to infer the position of the kinks in an analytically solvable model  -- the piecewise linear potential. Additionally, we demonstrate in the Appendices that our method allows to infer the number of metastabilities of models that are not susceptible to analytical investigations.  
It is important to stress that there are obvious limits to the method. For example, if two extremal points in the potential are in close proximity, or if the trap potential introduces a very rugged landscape, the sharpness of the response function scaling test is reduced.

Experimentally, we expect our method to be useful in analyzing transport of interacting particles through a narrow channel, and against a potential gradient. 
The robust experimental setup by Thorneywork et al.\cite{thorneywork2020direct} that recently designed colloid transport in the presence of multiple traps by an optical tweezers setup could be a potential avenue to validate our method. Within this setup, collecting measurements on the particle escape rates at different particle densities would allow us to  predict the number of local extrema of the potential gradient. We recall that for the initial conditions, one would have to consider the particles to be located well within the trap, never to be too close to the absorbing boundary. Additionally, our analysis would change dramatically if a macroscopic set of particles would be detached from the MEC, and repositioned around a local minimum of the potential. 
In an alternative setup, one could also consider tracking the current of a boundary driven setup with an additional bulk drive represented by the trap. Here, initial conditions become immaterial, as the steady state current is associated with the mean escape rate.  This setup and its implications will be discussed in a subsequent publication.  
It is important to emphasize that the response function approach becomes particularly useful when escape times become too large due to large energy barriers, at which the kinks in the data become apparent. Higher-order response functions enhance the peaks -- the precursors of the kinks --  allowing their identification at lower energy barriers. However, this improvement comes at the cost of increased data acquisition, as higher-order responses require measurements of escape rates over a denser range of densities. Since the determination of escape rates is itself experimentally challenging and prone to under-sampling \cite{bebon2023controlling,dieball2025precisely,zunke2022first}, the optimal order of the response function must be chosen with care, depending on the specific experimental conditions.

In this work, we have restricted the trap potential to $1D$. Nevertheless, we point out that the treatment can be extended to higher dimensions, as well as to multi-channel activation problems.
It would be interesting to consider also non-diffusive activation processes introducing interacting systems \cite{doyon2023ballistic,bernardin2024macroscopic}, as well as particles with soft interactions, going beyond  lattice gas models \cite{dandekar2023dynamical,santra2022gap}.        
Finally, it may be interesting to apply novel approaches to use the finite $\beta \Delta U$ scaling close to the critical points of the escape rate $\Phi_{\rm{MP}}$ in order to obtain better predictions of the behavior away from the criticality \cite{hathcock2024phase}.

\begin{acknowledgements}
The numerical calculations reported in
this work were carried out on the Nandadevi and Kamet cluster, which are maintained and supported by the Institute of Mathematical Science’s High-Performance Computing Center. AP gratefully acknowledges research support from the Department of Atomic Energy, Government of India via Soft Matter Apex projects. OS gratefully acknowledges useful discussions on the experimental aspects of the work with Yael Roichman and Felix Ritort.
\end{acknowledgements}

\section*{Data availability}
All the data supporting the findings of this study are included inside the article.

\section*{AUTHOR DECLARATIONS}
\subsection*{Conflict of Interest}
The authors have no conflicts to disclose.

\appendix

\section{Single particle escape problem
\label{SI:Single particle}
}

Let us revisit the Kramers setup leading to the Arrhenius law for a single particle. This case will help to benchmark the many-body escape problem we focus on in this work. Consider first an overdamped particle, located in a $1D$ potential trap of size $L$, and coupled to a thermal bath at temperature $T$. For convenience, we fix the trap to have a global minimum at $X=0$ with $U(0)=0$, imposing a reflecting boundary condition.  At $X=L$,  the potential has a global maximum with $U(L)=\Delta U$, where we impose an absorbing boundary condition. We are interested in the mean escape time, i.e the mean time it takes the particle to reach the absorbing boundary at $X=L$ for the first time. In this work, we focus on deep potential traps, i.e. $\beta \Delta U \gg 1$,  where the mean escape time is long, and large deviation theory is relevant.  It is then useful to study the probability that the particle did not escape  -- the survival probability $S(t)$. For times much larger than the diffusive time, the survival probability takes the large deviation form $S(t) \asymp \rm{e}^{-t \Phi_\text{SP}}$, where $\Phi_\text{SP}$ is the particle escape rate, or the inverse of the mean escape time. 

For $M$ non-interacting particles, and provided that the survival time of all the particles in the trap is still large, we expect the survival probability to be $S(t) \asymp \rm{e}^{-t M \Phi_{\rm{SP}}}$, where $\Phi_{\rm{SP}}$ is the single particle escape rate. Hence, at the heart of our analysis we assume that for interacting particles in a deep trap, the survival probability for all the particles in the trap is still given by  $S(t) \asymp \rm{e}^{-t\Phi_{\rm{MP}}}$, where $\Phi_{\rm{MP}}$ is the escape rate with $M$ interacting particles in the trap.

\section{The macroscopic fluctuation theory recap
\label{sec:MFT} }

\begin{figure}[h!]
    \centering
    \includegraphics[width=\linewidth]{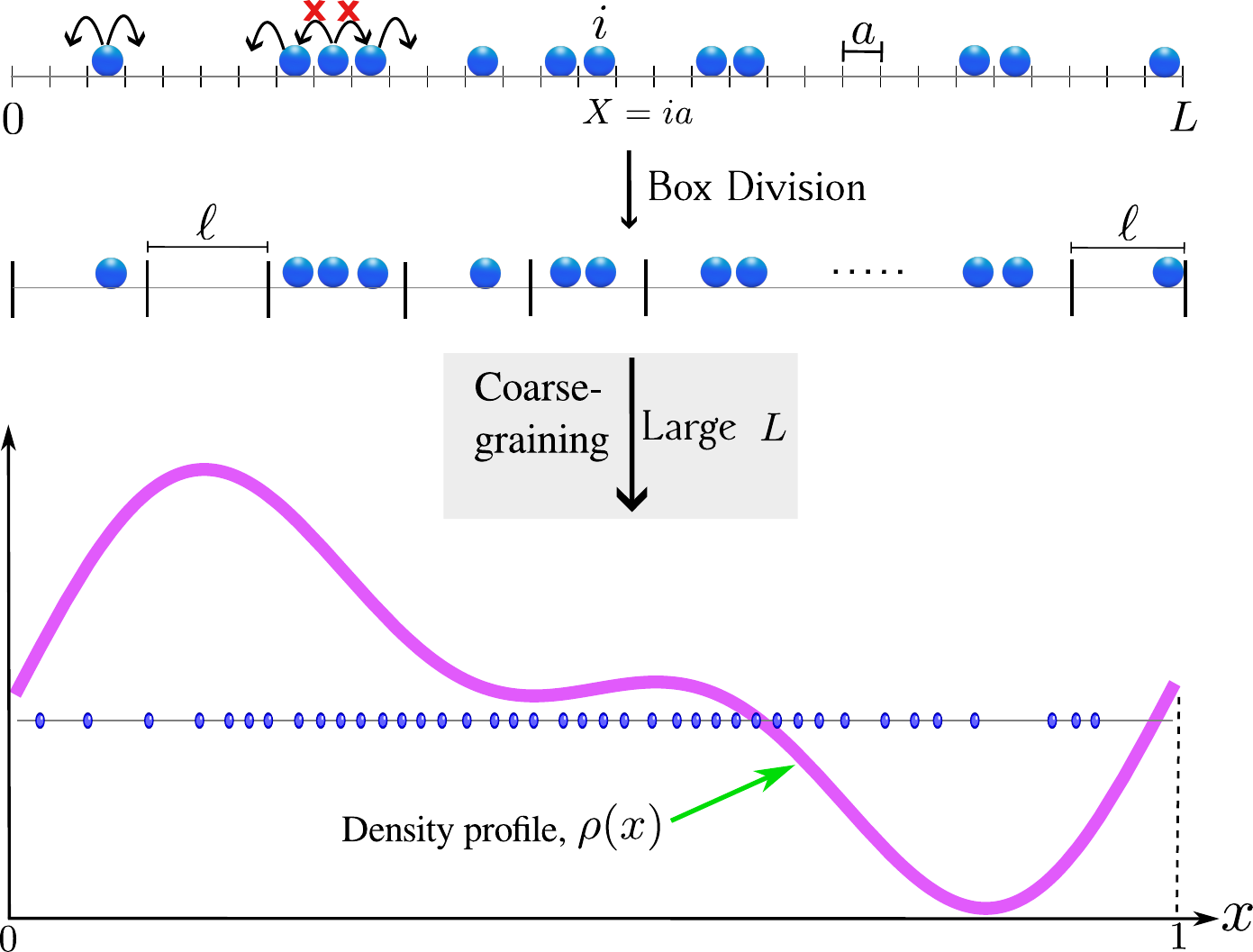}
    \caption{ The coarse-graining process: The top panel illustrates the microscopic dynamics of an exclusion process on a lattice. The middle panel depicts the process of division of the system into boxes of size $\ell$, which is necessary to define the local density(see Eq.~\eqref{Eq:Local density}). After coarse graining via diffusive rescaling, the system is described in terms of macroscopic density $\rho(x)$. The bottom panel shows a representative density profile.
}
    \label{fig:coarse graining}
\end{figure}

 The macroscopic fluctuation theory (MFT) is a hydrodynamic theory that deals with the nonequilibrium behavior of diffusive systems \cite{bertini2015macroscopic}. For example, consider an exclusion process on a lattice in 1-D as shown in top panel in  FIG.~\ref{fig:coarse graining}, where $a$ is the lattice spacing. The particle diameter is considered to be smaller than the lattice spacing $a$, and thus its information becomes irrelevant upon coarse-graining. In other words, the particle diameter does not influence the hydrodynamic description. The position of the $i^{th}$ lattice is $X=ia$. We assume $L$ to be large, so that a rescaled, dimensionless, continuous spatial coordinate $x=X/L$ could be introduced. Since the system is diffusive, we could also introduce a macroscopic time, $\tau=t/L^2$. The dimensions of other quantities transcend accordingly. (We note, however, that throughout the paper we set $a = 1$ for simplicity, thereby rendering all quantities effectively dimensionless.) Following the review by Derrida\cite{derrida2025lecture}, we can now define a macroscopic density. To this end, the system is partitioned into boxes of length $\ell$ (see middle panel of Fig.~\ref{fig:coarse graining}) such that $a \ll \ell \ll L$. The local macroscopic density $\rho(x,\tau)$ is then defined as the total number of particles contained within a box of size $\ell$, where the box begins at the lattice site with index $i = Lx / a$ and extends over a physical length $\ell$, divided by $\ell$. Mathematically,
\begin{equation}
\label{Eq:Local density}
    \rho(x,\tau)=\frac{1}{\ell}\sum_{i=\frac{Lx}{a}}^{\frac{Lx}{a}+\frac{\ell}{a}-1}n_i(t=L^2\tau),
\end{equation}
where, $n_i$ denotes the occupation variable at site $i$, taking the value $1$ if the site is occupied by a particle, and $0$ if it is vacant. Note that the dimension of the density is (length)$^{-1}$. Further, since there is no sink or source of particle in the  bulk, a continuity equation can be written in terms of macroscopic density and current.
 
At the hydrodynamic (large $L$) limit, the MFT captures the evolution of the particle density and current density,  $\Gamma=\lbrace \rho(x,\tau) ,j(x,\tau) \rbrace $. The dynamics of density and current density are encapsulated in the fluctuating hydrodynamics equations (restricted here to $1D$ and a single particle species for brevity)
\begin{subequations}
\label{eq:All fluc hydro eq}
\begin{align}
\label{eq:continuity equation}
\partial_\tau \rho & = -\partial_x j  \\
\label{eq:Fluc hydro}
j & = J(\rho)  + \sqrt{2\beta \chi/ L} \, \xi(x,\tau), \\  
   \label{eq:Fick law current}
    J(\rho) &= -D \partial_x \rho - \chi \beta  \partial_x U . 
\end{align}
\end{subequations}

Eq.~\eqref{eq:continuity equation} is the continuity equation, suggesting that particles can leave the system only through the boundaries.   Eq.~\eqref{eq:Fluc hydro} describes the current density using a Langevin equation \footnote{The Langevin equation has a weak noise, and therefore, time discretization questions are immaterial.}, where the deterministic part is given by Fick's law of the current $J$. The Fick's law current in \eqref{eq:Fick law current} depends on the diffusivity $D$ and mobility $\chi$, which are typically local functions of the density. Importantly, $D,\chi$ are experimentally measurable. The noise term in \eqref{eq:Fluc hydro} is weak as we consider a large system size $L$. $\xi(x,\tau)$ is a white noise term is both $x$ and $\tau$. The noise strength is found to depend on the mobility $\chi$ in order to satisfy the Gallavotti-Cohen fluctuation relations \cite{derrida2007non}. 

To our purposes, it is useful to recast the  Langevin description \eqref{eq:All fluc hydro eq} into a path probability  of $\Gamma = \lbrace \rho ,j\rbrace $ in a time window 
\begin{eqnarray}
    \label{eq:MFT fundamental}
    P\left[ \Gamma \right] & \asymp &\rm{e}^{- \frac{L}{\beta} \int^\tau _0  d\tau' \int^1 _0  dx \, \frac{\left( j - J(\rho)\right)^2}{4\chi} 
    }
\end{eqnarray}
The path probability \eqref{eq:MFT fundamental}, and thus observables are dominated by a saddle point solution (also known as the optimal fluctuation method), as MFT considers large systems  -- $L \gg 1$.

\section{The survival probability  in the MFT formalism
\label{sec:survival mft}}

To address the survival probability, it is important to state that within the MFT, we consider the trap to be occupied with mean density $\overline{\rho}=M/L$. Namely, we take the limits of $M,L$ to be large, but with a fixed ratio. 

Using the path probability \eqref{eq:MFT fundamental}, one can write the survival probability of all the particles in the trap $U$ using a conditional sum, where the paths are subjected to the boundary conditions 
\begin{equation}
\label{eq:boundary conditions}
    J(\rho)|_{x=0} =0 = \rho|_{x=1}, \quad \forall t,   
\end{equation}
and to the conservation of particle mass $\int dx \rho  = \overline{\rho} $. 
It was demonstrated that the optimal fluctuation dominating the survival probability consists of a time-independent density profile and a vanishing current density \cite{kumar2024arrhenius,kumar2024emerging}. Therefore, the survival probability can be expressed as the minimization problem 
\begin{eqnarray}
\label{eq:free minimization}
    S(t) &\asymp& \rm{e}^{- t \Phi_\text{MP} } , 
    \\ \nonumber 
    \Phi_\text{MP} &=& 
    \frac{L}{ \beta } \min_{\rho(x)} \int dx \, \frac{J^2}{4\chi}-\Lambda (\rho - \overline{\rho})  , 
\end{eqnarray}
and subjected to the boundary conditions \eqref{eq:boundary conditions}. Here $\Lambda$ is a Lagrange multiplier enforcing mass conservation in the trap.

Solving the minimization problem \eqref{eq:free minimization} together with the boundary conditions \eqref{eq:boundary conditions} boils down to solving a hard differential equation, obtained through an Euler-Lagrange formalism \cite{kumar2024arrhenius}. To recover eq.~(2) in the main text, 
it is useful to note that the deep trap condition $\beta \Delta U \gg 1$ was not yet invoked. Taking advantage of the deep trap condition, a perturbative method was developed \cite{kumar2024emerging}. It suggests rewriting the minimization problem using a useful transformation \cite{agranov2018narrow} \footnote{The importance of the transformation is by removing points where both the numerator and denominator of the integrand in \eqref{eq:free minimization}  vanish. }
\begin{equation}
\label{eq:F transformation}
F[\rho] = \int^\rho _0 dz \frac{D[z]}{2\sqrt{\chi[z]}} .      
\end{equation}
Thus, the minimization problem reads 
\begin{eqnarray}
\label{eq:F minimization}
    \Phi_{\rm{MP}} &=& \frac{L} { \beta}  \min_{F(x)} \int dx J^2 _F - \Lambda (\rho(F)-\overline{\rho} )
    \\ \nonumber 
    J_F &=& \partial_x F + \frac{1}{2}\chi^{1/2}\beta \partial_x U
    \\ \nonumber
    J_F |_{x=0} &=& F|_{x=1} =0.  
\end{eqnarray}
The optimal profile $F$, solving the minimization problem \eqref{eq:F minimization}  can be developed as a perturbative series, i.e. $F=F_{eq}+ \omega \delta F+O(\omega^2)$.  Here $F_{eq}$ is the equilibrium $F$ profile which follows $J_F (F_{eq}) = 0 $, and $\omega$ is a self consistent small parameter accounting for the perturbation $\delta F$. The equilibrium profile $F_{eq}$, as defined, violates the absorbing boundary condition in \eqref{eq:F minimization}. The violation is expected to be small in the deep trap approximation as the density of particles close to the peak of the potential is exponentially small with $\beta\Delta U$. 
It is therefore the perturbative correction that rectifies the absorbing boundary condition. This procedure boils down to setting  
\begin{equation}
\label{eq:omega condition}
\omega = F_{eq}|_{x=1}.  
\end{equation}

The self-consistent perturbation theory requires $\omega \ll 1$ when $\beta \Delta U \gg 1$. In this case, the escape rate is estimated asymptotically  $\Phi_{\rm{MP}} \asymp \omega^2 $. It should be stressed that the perturbative analysis does not replace the minimization problem in obtaining a full expression to the escape rate $\Phi_{\rm{MP}}$. Instead, $\omega$ captures the exponential part in $\Phi_{\rm{MP}}$. To successfully capture the pre-exponential of $\Phi_{\rm{MP}}$, one must revert back to  \eqref{eq:F minimization}.

\section{Survival probability for the piecewise linear potential 
\label{sec:SEP Upwl analysis}
}

Let us apply the formalism in Appendix \ref{sec:survival mft} in order to find the escape rate of SEP from the piecewise linear potential trap $U_{\rm{pwl}}$. Within the MFT, the SEP is characterized by $D=1$ and $\chi = \rho(1-\rho)$. The $F$ transformation \eqref{eq:F transformation} leads to $\rho  = \sin^2 F$. Thus one recovers $\omega^2 = \rm{e}^{-\beta \Delta U}A$, where $A$ can be recovered from 
\begin{equation}
\label{eq:Formal A}
    1- \overline{\rho} = \int dx \, \frac{1}{1+A \rm{e}^{-\beta U(x)}}.
\end{equation}
For the piecewise linear potential, \eqref{eq:Formal A} can be solved exactly, leading to
\begin{equation}
\label{eq:A polynomial}
    \rm{e}^{-2\beta \Delta U \overline{\rho} } = \frac{1+A k^3}{1+A} \left( 
    \frac{1+A k^2}{1+A k}
    \right)^3. 
\end{equation}

In the regime $0<\overline{\rho}<1/6$, we expect from the MEC that $A\asymp \rm{e}^{2\beta \Delta U \overline{\rho}}$, and so $Ak^2,Ak^3 \rightarrow 0$ in the large $\beta \Delta U$ limit, where we recall the definition $k = \rm{e}^{-\beta \Delta U /3 }$. Under these approximations, \eqref{eq:A polynomial} can be relaxed into 
\begin{equation}
    \rm{e}^{\beta \Delta U (2\overline{\rho}-1/3)} = b(1+b)^3, 
\end{equation}
where $b= Ak$. For $\overline{\rho}$ at a finite distance away from the critical density $1/6$,  $b$ is exponentially small in $\beta \Delta U$. Thus, one can relax $b(1+b)^3\approx b(1+3b)$, discarding cubic contributions from $b$. This results in the quadratic $b$ equation, with a single positive root, leading to  $\Phi_\text{MP} \asymp \rm{e}^{-\frac{2}{3}\beta \Delta U}b$, where 
\begin{equation}
    b= \frac{1}{6}\left( -1 + \sqrt{1+12 \rm{e}^{-2\delta \rho  \beta \Delta U}} \right)
\end{equation}
and $\delta \rho =1/6-\overline{\rho}.$ 
This reproduces eq.~(10) of the main text when  eq.~(8) in the main text is used to determine $\mathcal{F}$.

\begin{figure*}
    \centering
    \subfigure[]{
        \includegraphics[width=0.45\textwidth]{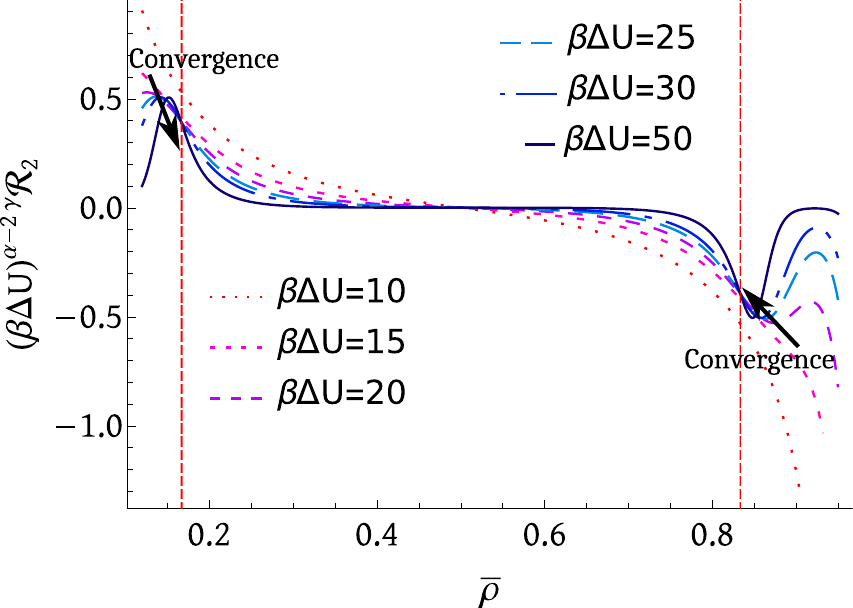}  
    }
    \hfill
    \subfigure[]{
        \includegraphics[width=0.45\textwidth]{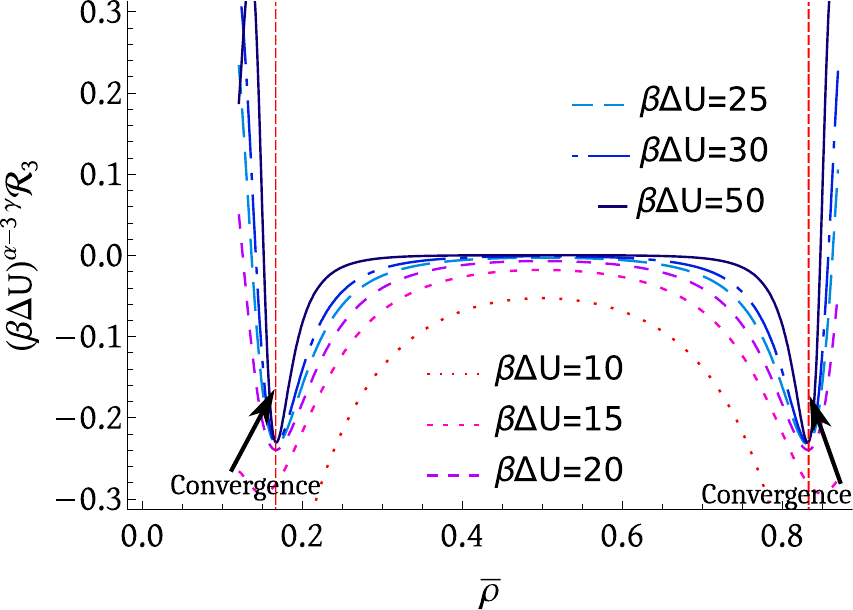}  
    }
    \caption{ The rescaled response functions $\mathcal{R}_n/(\beta\Delta U)^{n\gamma-\alpha}$ for piecewise linear potential $U_\text{pwl}$(see FIG. ~\ref{fig:Upwl and 1-g} of the main text) are shown for $n=2$ (panel a) \& $n=3$ (panel b) and for different $\beta\Delta U=\{10,15,20,25,30,50\}$, as indicated by the plot-legends. The response functions are obtained by first applying the perturbative approach outlined in \ref{sec:SEP Upwl analysis} and the main text. After computing $\Phi_\text{MP}$, we use Eq.~\ref{Free-energy-PT} of the main text to determine $\mathcal{R}_n$ for different $\beta\Delta U.$ The two vertical dashed red lines highlight the positions of the expected kinks, as determined by the MEC argument. The scaling behavior with $\alpha = \gamma =1 $ predicted from the theoretical analysis is confirmed. One can observe that the scaling convergence is reasonable starting from $\beta \Delta U = 15$, and thus the response functions are able to capture the locations of the kinks at finite $\beta\Delta U$ values. It is noteworthy that such high values of $\beta\Delta U$ are required due to the limitations of the perturbative approach. We expect real data to collapse even for lower $\beta \Delta U$ values. }
     \label{fig:Upwl Full Numerics}
\end{figure*}

So far, we have focused on an analytical analysis of the escape rate for the piecewise linear potential. Since we have employed approximation techniques, restricted the analysis to the range $0<\overline{\rho}<1/6$, and since it is likely that most potential cannot be analyzed analytically, we turn to present the numerical analysis of the piecewise linear potential. 
FIG.~\ref{fig:Upwl Full Numerics} presents the response functions corresponding to  the piecewise linear potential  over the entire range of $\overline{\rho}$. The numerical method outlined here is general and not limited to the case of the piecewise linear potential. It has also been utilized to compute the scaling functions for two additional potentials, as described in Appendix \ref{sec:More models}. The numerical steps are as follows :

\begin{enumerate}
    \item Fix $\beta \Delta U$ at some finite value.
    \item For this $\beta\Delta U$ value, an array of $A$ is constructed. This is the most technical step of the numerics, as the elements of the array must be carefully selected to cover the entire range of $\overline{\rho}$, from $0$ to $1$. It can be checked, using (\ref{eq:Formal A}), that selecting the elements of the array to range from $0$ to $e^{2\beta\Delta U}$ fulfils this requirement.
    \item For every element of the array of $A$, the corresponding $\overline{\rho}$ is computed using (\ref{eq:Formal A}) and the corresponding response function is computed using $\mathcal{F}=-\frac{1}{\beta \Delta U}\text{Log}\Big(\text{element of array of $A$} \times e^{-\beta\Delta U}\Big)$.
    \item At the end of step (3), we have $\mathcal{F}$ for different values of $\overline{\rho}$ ranging from $0$ to $1$. The second and the third derivatives of $\mathcal{F}$ is then computed numerically to yield $\mathcal{R}_2$ and $\mathcal{R}_3$ respectively.
    \item Steps (1) to (4) are then repeated to find $\mathcal{F}$, $\mathcal{R}_2$ and $\mathcal{R}_3$ for different values of $\beta\Delta U$.
\end{enumerate}

Note that in FIG.~\ref{fig:Upwl Full Numerics}, 
as $\beta\Delta U$ increases, the peaks in $\mathcal{R}_2$ and $\mathcal{R}_3$ progressively shift towards the true position of the kink in the function $g(\bar{\rho})$, as predicted by the MEC argument (also see FIG.~\ref{fig:Upwl and 1-g} in the main text). This is quite common in thermodynamic phase transitions as well. 

In the following section, we demonstrate that inferring the kinks in the function $g$ using response functions is not restricted to the case of piecewise linear potentials. Instead, we show that this approach is applicable to arbitrary potential landscapes. However, before proceeding, it is important to first discuss the experimental applicability of the method.

Experimentally, the escape rate $\Phi_\text{MP}$ of the rightmost particle can be measured for various particle densities $\overline{\rho}$. From the obtained values of $\Phi_\text{MP}$, the function $\mathcal{F}$ can be determined using eq.~(8) in the main text. Once $\mathcal{F}$ is known, the response functions $\mathcal{R}_2$ and $\mathcal{R}_3$ can be obtained by evaluating the second and the third order derivative respectively of $\mathcal{F}$ with respect to $\overline{\rho}$. These response functions will then reveal the kinks in the function $g$ through the peaks observed in the response functions.

\begin{figure*}[htbp]
    \centering
    \subfigure[]{
        \includegraphics[width=0.45\textwidth]{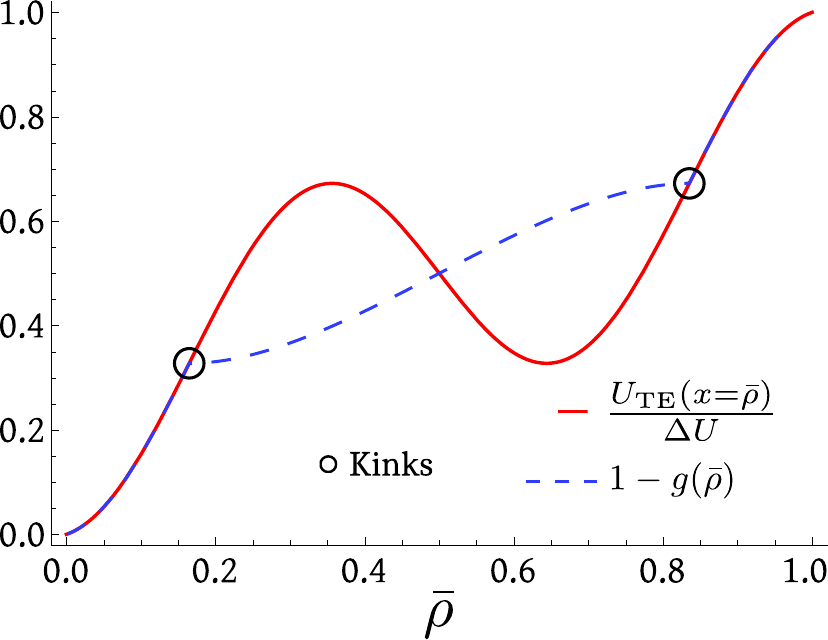}  
    }
    \hfill
    \subfigure[]{
        \includegraphics[width=0.45\textwidth]{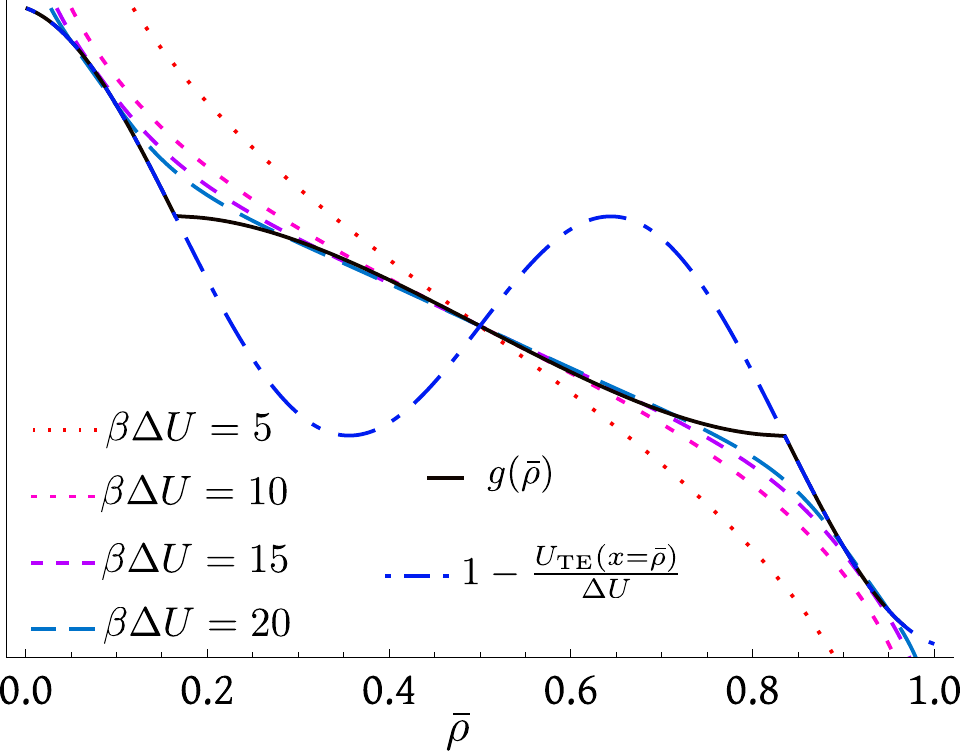}  
    }

    \vspace{0.5cm}  
    \subfigure[]{
        \includegraphics[width=0.45\textwidth]{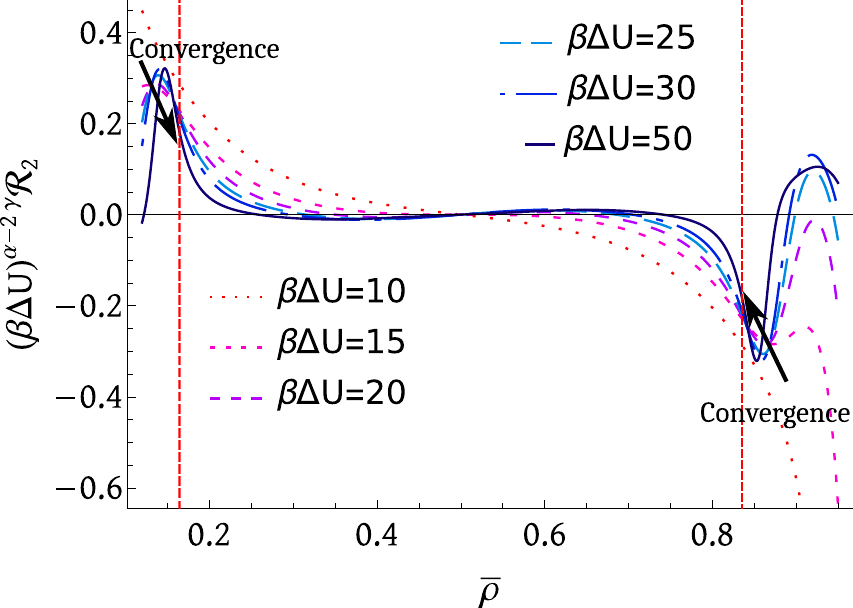}  
        }
\hfill
    \subfigure[]{
        \includegraphics[width=0.45\textwidth]{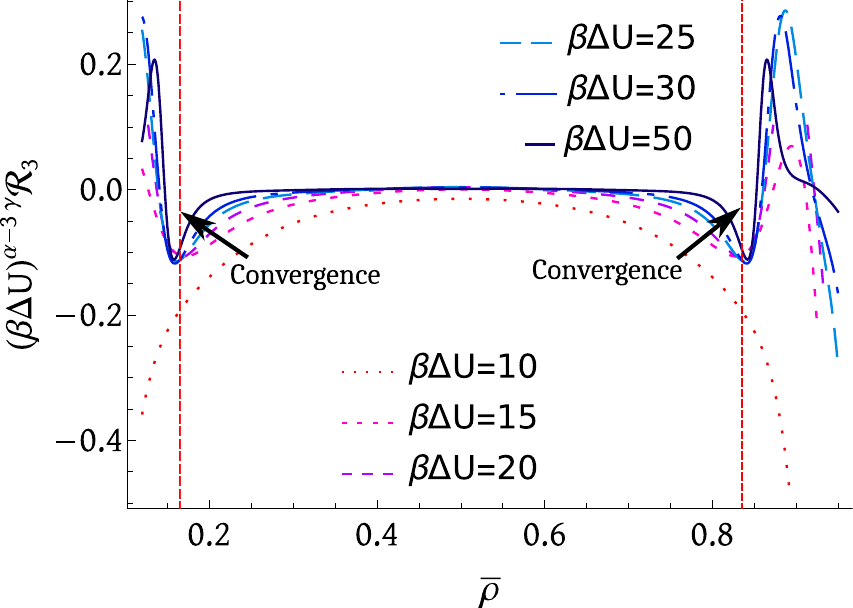}  
        }

    \caption{ Exploring metastabilities in a generic nonlinear potential landscape $U_{TE}/\Delta U  = \frac{1}{2}\left( x + \sin^2(3\pi x /2)\right)$, with two local extremal points. Panel (a): The potential $U_\text{TE}/\Delta U$ and the corresponding $1-g(\bar{\rho})$ are plotted as solid and dashed curves, respectively. The function $g$ is determined using the MEC argument, as detailed in the main text. Panel (b): The function $1-U_\text{TE}/\Delta U$ is represented by the dashed blue curve, while the function $g(\bar{\rho})$ is plotted as a solid line, as indicated in the legend. The remaining legend labels indicate the colors of the ``free energies'' curve for the values of $\beta\Delta U=\{5,10,15,20\}$. As $\beta\Delta U$ increases, the ``free energies'' $\mathcal{F}$ are seen to converge to the function $g(\bar{\rho})$. This behaviour is expected as $g(\bar{\rho})$ is obtained from the ``free energy'' $\mathcal{F}$ in $\beta\Delta U\to\infty$ limit. The kinks in $g(\bar{\rho})$ are not visible in the ``free energies'' curve for finite values of $\beta\Delta U$. They would become evident in the limit $\beta\Delta U\to \infty$ where $\mathcal{F}\to g(\bar{\rho})$. Panel (c \& d):  The rescaled response functions $\mathcal{R}_n / (\beta \Delta U)^{n\gamma-\alpha} $ are plotted for $n=2,3$  and $\beta \Delta U = \lbrace 10,15,20,25,30,50 \rbrace $, to reveal the convergence onto the actual positions of the kinks (indicated by vertical dashed red lines). Thus, the emergence of the kinks can be identified even for finite values of $\beta\Delta U$. The scaling exponents are $\alpha =1,  \gamma =1.15$. It is important to emphasize, however, that the exact values of $\alpha, \gamma$ are not necessary for identifying the emergence of the kinks. Due to the limitations of the perturbative approach, the divergences are detected only for values of $\beta\Delta U=15$ or higher. However, data obtained from experimental measurements is not expected to exhibit this limitation. Notice that away from the kinks, a  faster convergence is observed, as expected.}
    \label{fig:TE potential Numerics}
\end{figure*}

\section{Inferring metastable states in nonlinear potential traps \label{sec:More models}}

To demonstrate the predictive power of our theory, and potential pitfalls, we turn to analyze two general trap potentials that cannot be handled analytically. Such potentials appear ubiquitously in various biological settings as mentioned in the main text. In the first case, the potential is given by  
\begin{align}
    U_{\rm{TE}}/\Delta U = \frac{1}{2} (x + \sin^2 (3 \pi x /2 )),
\end{align}
 which exhibits two kinks in $g(\bar{\rho})$ according to the MEC. In the second case, we take the following form of the potential  
 \begin{align}
U_{XS} / \Delta U = \frac{1}{8}( 6 x  + 2 \sin^2 (5 \pi x /2 )),     
 \end{align}
 which features $4$ kinks in $g(\bar{\rho})$ according to the MEC. We demonstrate that the response functions $\mathcal{R}_2$ and $\mathcal{R}_3$ successfully capture these kinks in both cases.

\subsection{Inferring kinks in $g(\bar{\rho})$ for $\mathbf{U_{\rm{TE}}/\Delta U = \frac{1}{2} (x + \sin^2 (3 \pi x /2 ))}$}
\label{Two Kinks analysis}

Let us first consider the trap potential $U_{\rm{TE}}$. All the results in this section are obtained numerically. 
According to the MEC, $g$ exhibits two kinks. See   FIG.~\ref{fig:TE potential Numerics}(a). Yet, the emergence of the kinks cannot be identified directly, even for relatively high values of $\beta \Delta U$. 
 The response functions $\mathcal{R}_2$ and $\mathcal{R}_3$ capture the emerging kinks in  $g$ at finite $\beta \Delta U$. Here we have numerically computed the response functions using the algorithm of Appendix \ref{sec:SEP Upwl analysis}. The resulting function $\mathcal{F}$ and the response functions $\mathcal{R}_2$ and $\mathcal{R}_3$ are plotted in FIG.~\ref{fig:TE potential Numerics} for several values of $\beta\Delta U$. As $\beta\Delta U$ is increased, the peaks in the response function shift progressively closer towards the expected position of the kinks, predicted by the MEC argument.

\subsection{Inferring kinks in $g(\bar{\rho})$ for $\mathbf{U_{XS} / \Delta U = \frac{1}{8}( 6 x  + 2 \sin^2 (5 \pi x /2 )}$}
\label{Four Kinks analysis}

\begin{figure*}[htbp]
    \centering
    \subfigure[]{
        \includegraphics[width=0.45\textwidth]{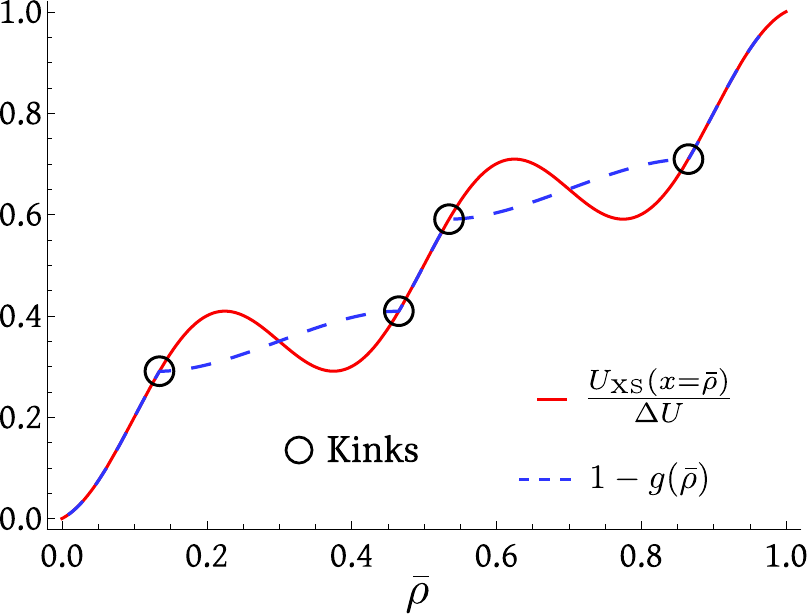}  
        \label{fig:fig1}
    }
    \hfill
    \subfigure[]{
        \includegraphics[width=0.45\textwidth]{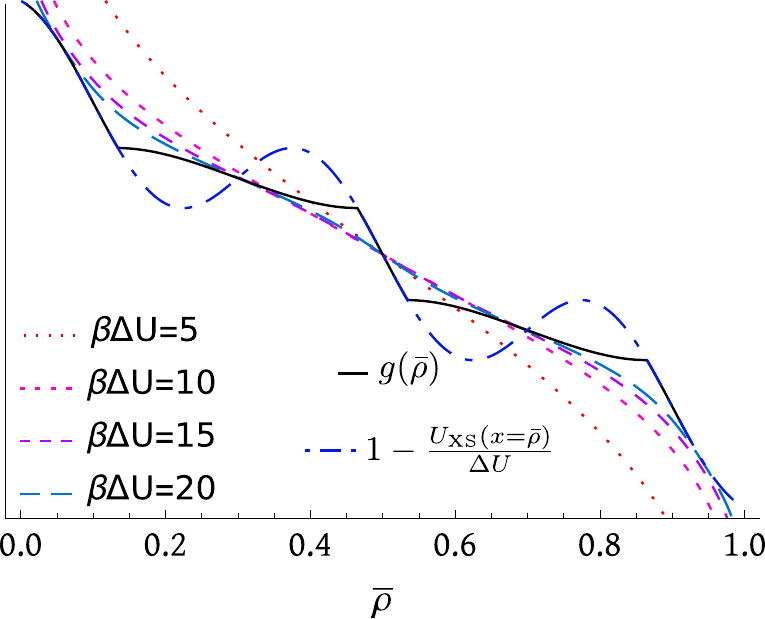}  
        \label{fig:fig2}
    }

    \vspace{0.5cm}  
    \subfigure[]{
        \includegraphics[width=0.45\textwidth]{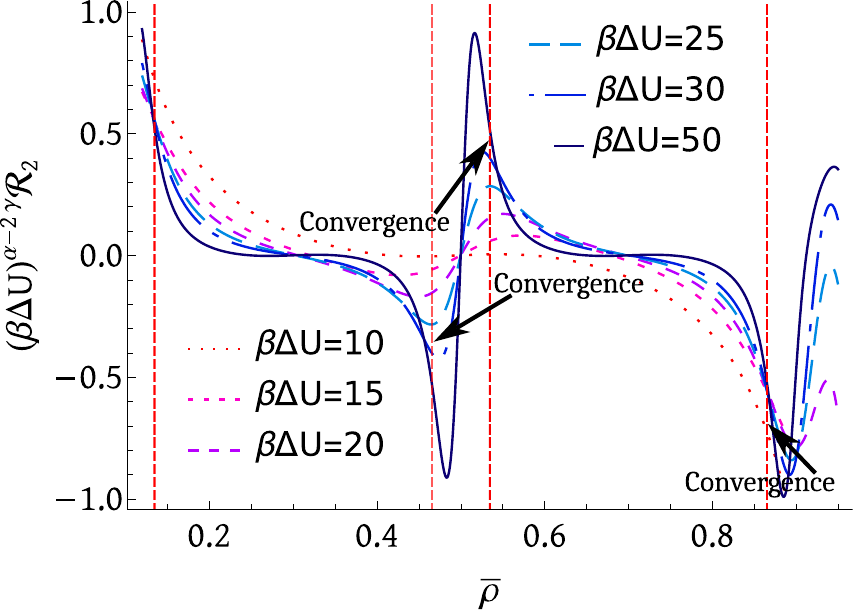}  
        \label{fig:fig3}
    }
    \hfill
    \subfigure[]{
        \includegraphics[width=0.45\textwidth]{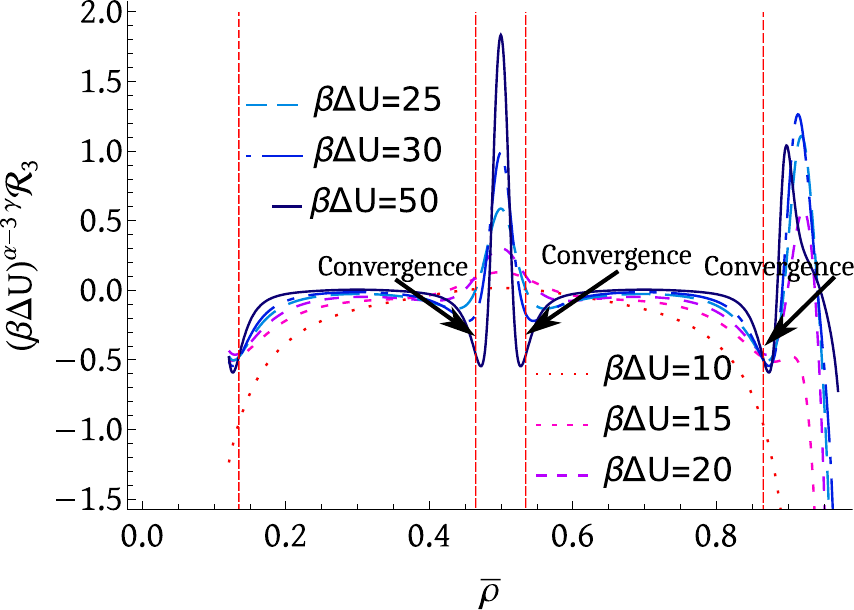}  
    }

    \caption{Exploring metastabilities in a potential landscape $U_{XS}/\Delta U=\frac{1}{8}(6x+2\sin^2(5\pi x/2))$, with four local extremal points. Panel (a): The potential $U_\text{XS}/\Delta U$ and $1-g$ are plotted as solid and dashed lines, with the actual positions of the kinks indicated. The locations of these kinks in $g$ are determined using the MEC argument, which predicts four kinks, each corresponding to a local extremum of the potential. Panel (b): 
    The functions $1-U_\text{XS}/\Delta U$ and $g$ are plotted as dashed blue curve and a solid line, respectively, as indicated in the legend. The other legend labels represent the ``free energies'' curve for $\mathcal{F}$ for various values of $\beta\Delta U=\{5,10,15,20\}$.  The kinks would become apparent only in the experimently unfeasible limit $\beta\Delta U \to \infty$, in which case $\mathcal{F}\to g$.  Panel (c\& d): The rescaled response functions $\mathcal{R}_n/(\beta\Delta U)^{n\gamma-\alpha}$ are plotted to infer the locations of the kinks at finite $\beta\Delta U$ values. These response functions are plotted for $n=2, 3$ for different values of $\beta\Delta U=\{10,15,20,25,30,50\}$, as indicated by the legends. The scaling exponents are $\gamma=1$, $\alpha=1$.
    Here, while it is evident that the diverging peaks of the response functions accurately pinpoint the actual positions of the kinks (denoted by vertical dashed red lines), the scaling is not that obvious. Nevertheless, determining the scaling exponents is not crucial for accurately identifying the kink positions.}
    \label{fig:Uxs potential }
\end{figure*}

In the present case, the function $g$, determined using the MEC argument, features four kinks, as shown in FIG.~\ref{fig:Uxs potential }. Following a similar approach as outlined in \ref{Two Kinks analysis}, the functions $\mathcal{F}$ and the response functions $\mathcal{R}_2$ and $\mathcal{R}_3$ are computed, which is shown in FIG.~\ref{fig:Uxs potential }. It is evident from the figure that, at finite values of $\beta\Delta U$, the response function $\mathcal{R}_3$ successfully identifies the emergence of the four kinks, as well as their positions. However, $\mathcal{R}_2$ fails to detect the first kink, although it is still able to identify the emergence of the other three kinks. The reason for this failure stems from the perturbative approach employed in the reference\cite{kumar2024emerging}, which forms the basis for the current analysis, assumes the limit $\beta\Delta U\overline{\rho}\gg 1$. In the case of the first kink, for which $\overline{\rho}$ is quite small, this approximation breaks down. Nevertheless, by increasing the value of $\beta\Delta U$ such that $\beta\Delta U\overline{\rho}\gg1$, the emergence of the first kink would also be identifiable in $\mathcal{R}_2$. It is important to note here that this is a limitation imposed by the perturbation theory. We do not expect such limitations if the escape rate data were obtained experimentally or through the direct simulation of the SEP.

\bibliography{alinference}

\end{document}